\documentclass{article}
\usepackage{times,cite}
\usepackage{graphicx,epsfig}    
\usepackage[tbtags]{amsmath}
\usepackage{longtable}
\usepackage[left=19mm,right=15mm,top=27mm,columnsep=15pt]{geometry}
\usepackage{subcaption}
\usepackage{hyperref} 
\usepackage{multirow}
\usepackage{authblk}
\usepackage{appendix}
\usepackage{xcolor}

\title{Simultaneous study of scattering and fusion hindrance near Coulomb barrier in $F+Pb$ systems}

\author[1,2]{Kamala Kanta Jena}
\author[3]{Bidhubhusan Sahu}
\author[4]{Jajati K. Nayak,}
\author[5]{Raj Preethi P}
\author[5]{B. K. Sharma}
\author[2]{Santosh Kumar Agarwalla}

\affil[1]{P. G. Department of Physics, Bhadrak Autonomous College, Bhadrak-756100, India}
\affil[2]{P. G. Department of Physics and Ballistics, Fakir Mohan University, Balasore-756019, India}
\affil[3]{School of Applied Sciences, KIIT Deemed to be University, Bhubaneswar-751024, India}
\affil[4]{Variable Energy Cyclotron Centre, 1/AF, Bidhan Nagar, Kolkata-700064, India}
\affil[5]{Department of Sciences, Amrita Vishwa Vidyapeetham, Ettimadai, Coimbatore-641105, India}

\begin{document}
\maketitle

\begin{center}
\end{center}

\abstract{%
A phenomenological optical potential is used to study the elastic angular distributions for the 
system $^{19}F+^{208}Pb$ close to the Coulomb barrier. This potential is constructed by taking into account the flexible potential developed by Ginocchio. The fluctuations in the real and imaginary parts 
of the optical model potential follow the trends of the threshold anomaly. The set of optical 
potential parameters needed to analyze the fusion cross sections of the same system are obtained 
through analysis of the scattering cross sections. Theoretical fusion cross-sections and results 
from four different experimental groups well agree for a range of energies. Several 
Fluorine (F) isotopes are used as projectiles in this study of fusion cross-sections by slightly 
altering the radial parameter. It was found that the fusion process occurs unfettered in 
the $^{19}F +^{208}Pb$ system below the Coulomb 
barrier but is seriously hindered in the case of its isotopic projectiles. \\

}

\section{INTRODUCTION}
 To examine the various nuclear characteristics, the analysis of experimental data from nucleus-nucleus scatterings using an optical model has been proven successful. In optical model analysis, phenomenological nuclear potentials like Woods-Saxon (WS), Gaussian, modified WS, and many more are employed. In heavy-ion elastic scatterings, the fluctuation in the real and imaginary components of an optical potential is seen as a crucial characteristic near the Coulomb barrier. Threshold anomaly (TA) is an interesting phenomenon, in the case of systems having heavy projectiles, in which the real component of the potential is practically constant at higher energies but rapidly increases as the incident energy gets closer to the Coulomb barrier. When the incident energy is below the barrier, it slowly starts to decline after reaching its maximum at the barrier. Around the barrier, thus, the variation takes a bell- shape. The imaginary part, on the other hand, shows nearly a constant magnitude at higher energies but decreases to a low value \cite{ref.1,ref.2,ref.3,ref.4,ref.5,ref.6,ref.7,ref.8,ref.9,ref.10} around the barrier in the same vicinity. In other words, when the collision energy rises above the top of the Coulomb barrier, the strength of imaginary potential rises rapidly and then its value becomes nearly constant. The maximum value of the real part can be two fold of the constant value it assumes at higher energies \cite{ref.11}. This anomalous variation is caused by the coupling of several elastic and quasi-elastic response channels. This is explicable by the dispersion relation developed by Byron and Fuller \cite{ref.12} using the causality principle. This study demonstrates how the optical potential, which has an imaginary component that is noticeably small, corresponds with TA occurrences and explains the fusion cross-section. In our discussions, we take into consideration a semi-classical heavy-ion elastic collision system, $^{19}F + ^{208}Pb$, whose experimental results may be interpreted in terms of an optical model by employing a complex potential with the appropriate parameterization. Lin et al. \cite{ref.13} carried out experimental measurements and theoretical analyses of the system. The angular distributions were observed with a fluorine beam ($^{19}$F) at six energies ranging from 80.6 MeV to 93.5 MeV in the center-of-mass frame. We analyze the outcomes for the same energy range to broaden our investigation of elastic scattering with a focus on TA. For TA analysis, mostly spherical nuclei have been investigated \cite{ref.14}. We work on one that is deformed. A sizable static $^{19}F$ nucleus with deformation is present in the system $^{19}F + ^{208}Pb$ \cite{ref.15}. To explore the nuclear characteristics, $^{19}$F has also been utilized as a projectile 
 \cite{ref.16,ref.17,ref.18,ref.19,ref.20} for decades in several elastic scatterings with various targets and incidence energies. All these facts motivate our team to choose the projectile $^{19}$F and realize the versatility of our potential. \\
 
We use a phenomenological optical potential \cite{ref.21,ref.22} based on a short-ranged, smooth, and analytically solvable asymmetric potential developed by Ginocchio \cite{ref.23} that possesses the versatility to control the volume and surface regions of the potential. The parameters dealt in the optical potential are significantly less in number. The experimental results of $^{16}O+^{28}Si$ and $^{12}C+^{24}Mg$ systems were fairly explained by G. S. Mallick et al. \cite{ref.21} over a wide range of energy by using this potential. The interesting feature of our potential is the neck structure near the Coulomb barrier. This non-trivial feature helps us match theoretical data with experimental data. The potential agrees with the presence of ‘threshold anomaly’ because of the fast rise of the imaginary part with the rapid fall of the real part of the potential as the incident energy rises above the Coulomb barrier. The optical potentials used by most of the researchers deal with large imaginary parts. The absorption of a major share of partial waves cannot be avoided in cases having large imaginary parts. The imaginary parts remain below 12\% of their corresponding real parts in Ref.\cite{ref.13} for incident energies 80.6 – 85.2 MeV but exceed 29\% for 87.9 – 93.5 MeV. The fact that high imaginary parts substantially destroy the resonance states generated by the volume part of the effective potential cannot be ruled out. Hence, a small imaginary part may be more convincing. In this work, we use the potential where the imaginary part is very small as compared to its real part.\\

We extend the applicability of our optical potential in fusion as well. We use the potential to analyze fusion cross-section data obtained from four different experiments performed by D. J. Hinde et al. \cite{ref.24}, B. B. Back et al. \cite{ref.25}, K. E. Rehm et al. \cite{ref.26} and Zhang Huanqiao et al. \cite{ref.27} for the same collision system $^{19}F + ^{208}Pb$ but over different energy ranges. The analysis of fusion cross-sections involves mostly the same set of parameters used for the analysis of elastic scattering cross-sections. \\

As far as fusion hindrance at sub-barrier energies for drip-line nuclei is considered, the periodic table expanded (to now include 118 elements), and super heavy elements (SHE) were made available to humankind due to conceptual and experimental developments in physics. According to conventional content, elements with more than 104 protons shouldn't exist since the element would undergo spontaneous fission if the fission barrier had been zero. Yet again, the stabilization of these elements and the formation of SHEs with distinct features are caused by quantum shell effects. Even though the fusion process between massive nuclei has been well studied thus far, the fusion probability between massive nuclei is dependent on the charge product $Z_PZ_T$ of the projectile and the target. This is because when the charge product grows, the Coulomb repulsion between them grows, decreasing the likelihood of fusion. Nevertheless, nuclear processes involving heavy ions are utilized to produce SHEs. The formation of SHEs involves the use of both cold fusion and hot fusion nuclear processes. \\

The doubly magic nucleus $^{208}Pb$ is employed as a target in cold fusion reactions together with the suitable projectile. The one-dimensional barrier penetration model, which takes into consideration the coupling of inelastic excitations, has been noted to accurately represent the fusion cross section for charge products smaller than 1800. On the other hand, in contrast to the model's calculated results, the fusion cross section is hampered when the charge product is greater than 1800 \cite{ref.28}. But in addition to the charge product, other factors that affect nuclear fusion between heavy nuclei include the nuclear structures of the projectile and target. According to reports, the fusion probability is significantly influenced by the number of valence nucleons outside of a major shell closure \cite{ref.28,ref.29}. In the fusion processes, $^{130}Xe+^{86}Kr$ and $^{136}Xe+^{86}Kr$, where the nucleus $^{136}Xe$ has a closed neutron shell N=82 and the neutron number of the nucleus $^{130}Xe$ is 76, six neutrons fewer than the closed shell, the evaporation residue cross sections were determined by Oganessian et al. \cite{ref.30} in 1987. They discovered that, in the vicinity of the Coulomb barrier, the measured evaporation residue cross sections for the fusion process $^{136}Xe+^{86}Kr$ are about two to three orders of magnitude greater than those for the fusion reaction $^{130}Xe+^{86}Kr$. The enhancement of the evaporation residue cross sections near the Coulomb barrier region between the double closed shell nuclei $^{208}Pb$ and $^{48}Ca$ is also pointed out by Oganessian et al. \cite{ref.31} in 2001. The dependence of fusion on the nuclear shell structure was investigated by K. Satou et al. \cite{ref.32} in 2002 for the two reaction systems $^{82}Se+^{138}Ba$ and $^{82}Se +^{134}Ba$, where the nucleus $^{138}Ba$ has a closed neutron shell N=82 while the nucleus $^{134}Ba$ has a neutron number N=78; four neutrons less than the closed shell. The fusion reaction $^{82}Se+^{138}Ba$ takes place without hindrance, but $^{82}Se+^{134}Ba$ fusion is significantly hindered, as is typically observed in major reaction systems with the charge product $Z_PZ_T$ $\ge$ 1800 of the projectile and target. These results suggest that a crucial part of the low-energy fusion process involves the shell structure. We analyze the isotopic dependence in the $^{19-23}F+^{208}Pb$ systems to realize how the nuclear shell structure affects the fusion process. \\

 The paper is organized to discuss the formulation of the optical model based on Ginocchio potential in section $\bf {2}$. Section $\bf {3}$ explains the application of our optical potential to the elastic scattering of tightly bound projectile $^{19}F$ by $^{208}Pb$ target at energies near the Coulomb barrier and TA phenomena thereof. The fusion cross-sections for this system are also presented and compared with the data from the various experiments. Finally, the summary and conclusions are presented in section $\bf {4}$.

\begin{figure}
\begin{center}
\includegraphics[width=10.0cm]{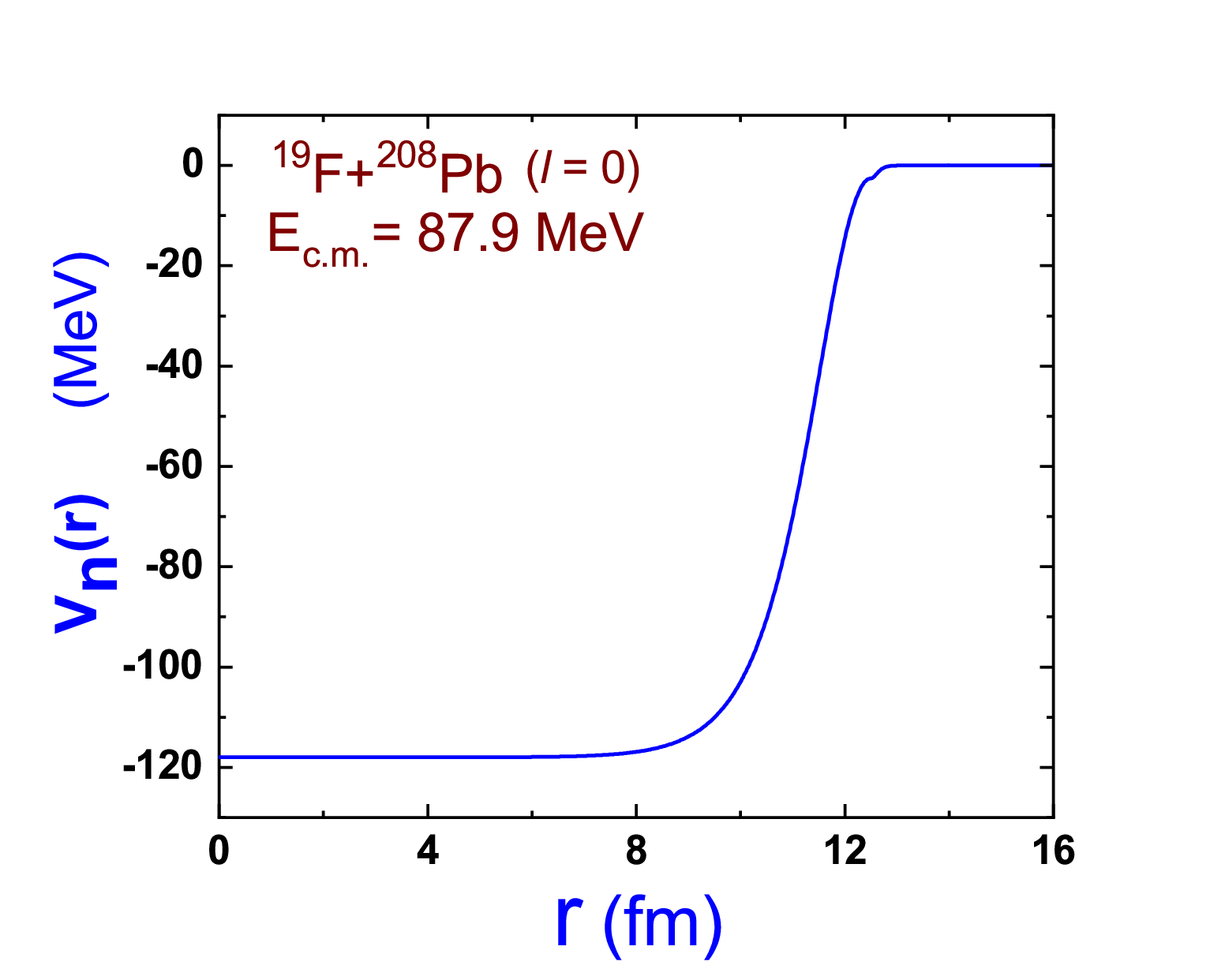}
\caption{%
The plot of real part $V_n(r)$ of optical potential for the system $^{19}F + ^{208}Pb$ at incident energy $E_{c.m.}$ = 87.9 MeV. Values of parameters are taken to be $R_0$ = 12.5 fm, $B_0$ = 118 MeV, $B_1$ = 7.0, $B_2$ = 0.6 and $V_B$ = 2.6 MeV. The curve shows a neck formation at $R_0$.
}
\label{fig1}
\end{center}
\end{figure}
\begin{figure}
\begin{center}
\includegraphics[width=10.0cm]{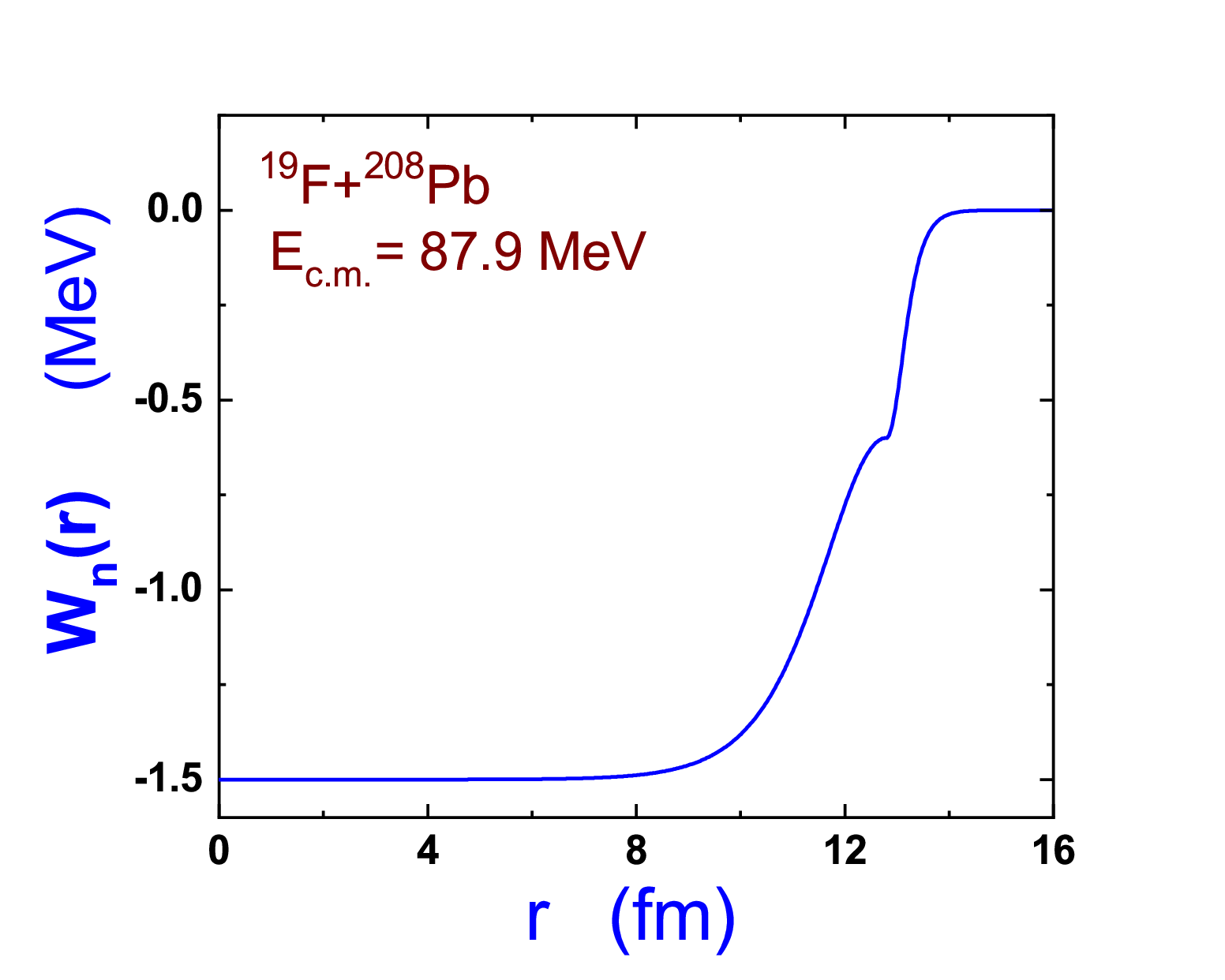}[t]
\caption{%
Imaginary part $W_n(r)$ of optical potential for the system $^{19}F + ^{208}Pb$ at incident energy $E_{c.m}$= 87.9 MeV. Values of parameters are $R_{0W}$ = 12.8 fm, $W_0$= 1.5 MeV, $V_{BW}$ = 0.6 MeV, $W_1$ = 1.4 and $W_2$= 0.001.
}
\label{fig2}
\end{center}
\end{figure}
\section{FORMULATION OF THEORY}

The phenomenological optical potential used here is based on a short-ranged, smooth, and analytically solvable asymmetric potential developed by Ginocchio \cite{ref.23} and used by others \cite{ref.21, ref.22, ref.33, ref.34, ref.35} which possesses the versatility to control the volume and surface regions. A potential to describe nucleus-nucleus interaction usually consists of Coulomb potential $V_C(r)$ due to electric charges of two nuclei and nuclear potential $V_N(r)$. Taking the centrifugal force into account, the effective potential $V_{eff}(r)$) for nucleus-nucleus collision with reduced mass $\mu$ and orbital quantum number $l$ can be described by Eq.1 in which the last term represents the potential owing to centrifugal force.
\begin{equation}
V_{eff}(r) =  V_N (r) + V_C(r) + \frac{l(l+1){\hbar}^2}{2\mu r^2}
\end{equation}
The nuclear part $V_N(r)$ is an optical potential. It is taken from Ref.\cite{ref.23} by considering the value of parameter $\lambda$=1, where $\lambda$ parameter is responsible for the flatness of the potential. Nuclear potential $V_N(r)$ is the most important part of the effective potential which is not uniquely described to date. But $V_N(r)$, as has been argued in many articles, takes a complex form to describe the experimental observations. We also consider $V_N(r)$ to be complex and represent as, $V_N(r)$ = $V_n(r) + iW_n(r)$. The variable $\mu$ represents the reduced mass of the projectile-target system and is defined as $\mu= \frac{m_P \times m_T}{m_P+m_T}$, where $m_P$ and $m_T$ being the masses of projectile and target respectively. \\

Following the potential developed in Ref.\cite{ref.23}, which is further simplified in Ref.\cite{ref.22}, we consider the real part $V_n(r)$ of the potential as given in Eq.2 by putting $\lambda$=1.
\begin{equation}
V_n(r)= \left\{ \begin{array}{ll}
           \frac{-V_B}{B_1}\left[ B_0+(B_1-B_0)(1-y_1^2) \right]  &  \textrm{if}~~ 0<r<R_0\\
           & \\
           \frac{-V_B}{B_2} \left[ B_2(1-y_2^2)\right] & \textrm{if}~~r\ge R_0\\
          \end{array}
 \right.
 \label{eq_nuclpot}
\end{equation}
On substitutions of $y=tanh\rho_n$, $\rho_n=(r-R_0)b_n$, and $V_B = V_{01} B_1$ = $V_{02}B_2$, we find 
\begin{equation}
V_n(r)= \left\{ \begin{array}{ll}
           -V_{01}\left[B_0+\frac{(B_1-B_0)}{cosh^2\rho_1} \right]  &  \textrm{if}~~ 0<r<R_0\\
           & \\
           -V_{02}\left[\frac{B_2}{cosh^2\rho_2} \right] & \textrm{if}~~r\ge R_0\\
          \end{array}
 \right.
 \label{eq_nuclpot2}
\end{equation}
Here, $V_B$ is the height of the barrier. The slope parameter $b_n$ is given by, $b_n=\frac{\sqrt{2 \mu V_B}}{\hbar^2B_n}$ in which n = 1 or 2. The radial distance $R_0$ in the surface region is close to the radial position of the effective S-wave barrier potential. The depth of potential at origin and $R_0$ are controlled by the parameters $B_0$ and $V_B$ respectively. Slope parameter $b_n$ on either side of $R_0$ depends on $B_n$ and $V_B$. Parameters $B_0$ and $B_1$ specify the potential for $r\leq R_0$. $V_{01}$ specifies the strength in that region, and is given by,

$V_{01} = \frac{V_B}{\lambda_1^2 B_1 + \frac{1-\lambda_1^2}{2}} = \frac{V_B}{B_1}$ for $\lambda _1$ =1. 
The parameter $\lambda _1$  controls the flatness of the potential in the region for $r\leq R_0$. Similarly, the parameters $B_2$ and $V_B$ specify the potential for the region $r > R_0$, where $R_0$ is sum of radii of two interacting nuclei, i.e., $R_0{= r_0(A_P^{\frac{1}{3}} + A_T^{\frac{1}{3}})}$= $R_1$ + $R_2$. The quantities $\rho_1$ and $\rho_2$ are the transformed distance variables and given by, $\rho_n = (r-R_0)\frac{\sqrt{2mV_B}}{\hbar^2B_n}= (r-R_0) b_n$.
With the above consideration, the real part $V_n$(r) of the optical potential for the collision system $^{19}F + ^{208}Pb$ is depicted in Fig.1. The potential has two regions; volume and surface. Two parts of the potential corresponding to the volume region and surface region are connected at $r = R_0$ satisfying the analytic continuity.
Unlike monotonous fall with ‘r’ in a nuclear potential of a standard Woods-Saxon form, our optical potential shows a neck-formation near $r = R_0$. The optical potential consists of two analytically solvable regions, namely, the volume region and the surface region. The regions are smoothly joined near r = $R_0$ forming a neck-like structure. We refer the location to an analytic junction \cite{ref.34, ref.35}, where the two regions of the potential meet each other. As the name suggests, the junction is analytically solvable and the Schrodinger equation can be solved there. The structure appears unusual, but our consideration ensures indifference in two parts of potential and keeps the respective derivatives (concerning ‘r’) the same at the meeting point satisfying analytic continuity. This new feature helps us suitably explain the differential scattering cross-sections and fusion cross-sections over a wide range of energies. The feature also enables us to apprehend the effects of frictional forces, resonance in the formation of a composite binuclear system and transfer of one or cluster of nucleons from the target to the projectile and/or vice versa in this configuration, when the bombarding nuclei touch each other in the surface region around $r = R_0$. \\

The form of the imaginary part $W_n(r)$ is similar to that of the real part, but its strength differs. The imaginary part in strength is weaker than that of the real part, i.e., the real part with a larger value is very deep and the imaginary part with a smaller value is comparatively weak. With substitution $V_{0nW}=\frac{V_{BW}}{W_n}$, the imaginary part is given by Eq.4 and its behaviour is plotted in Fig.2 with a suitable set of parameters.
\begin{equation}
W_n(r)= \left\{ \begin{array}{ll}
            -V_{01W}\left[W_0+\frac{(W_1-W_0)}{cosh^2\rho_1} \right]  &  \textrm{if}~~ 0<r<R_{0W}\\
            & \\
            -V_{02W}\left[\frac{W_2}{cosh^2\rho_2} \right] & \textrm{if}~~r\ge R_{0W}\\
           \end{array}
  \right.
  \label{eq_nuclpot3}
 \end{equation}
 The parameter $W_0$ represents the depth of imaginary potential at the origin and $V_{BW}$ controls the depth of potential at $R_{0W}$. The other two parameters, namely, $W_1$ and $W_2$ are slope parameters. Parameter $W_1$ specifies the potential for $r \leq R_{0W}$, whereas, $W_2$ specifies the potential for $r > R_{0W}$. We use a set of these parameters to represent imaginary part in Fig.2. The Coulomb potential for the projectile nucleus and target nucleus interacting system is given by Eq.5 as follows.
 \begin{equation}
V_C(r)= \left\{ \begin{array}{ll}
           \frac{Z_p Z_T e^2}{2R_C^3}(3R_C^2-r^2)  &  \textrm{if}~~ r<R_C\\
           & \\
           \frac{Z_p Z_T e^2}{r} & \textrm{if}~~r > R_C\\
          \end{array}
 \right.
 \label{eq_coulpot}
\end{equation}
Here, $R_C = r_C (A_P^{1/3} + A_T^{1/3})$;$A_P$ and $A_T$ being the mass numbers of projectile and target nuclei respectively. $Z_P$ and $Z_T$ are the atomic numbers of those nuclei. The value of the Coulomb radius parameter $r_C$ is taken to be 1.33 fm. Neglecting the centrifugal term with orbital quantum number $l$ = 0, Eq.1 describes effective potential as :
\begin{equation}
 V_{eff}(r)=V_n(r)+iW_n(r) +V_C(r)
\end{equation}
The real part of the effective potential $V_{eff}(r)$ is depicted in Fig.3 with a set of parameters considered earlier for the real part in Fig.1. With the above effective potential for the various partial waves ($ l $), we solve the following Schrodinger equation to obtain the total scattering amplitude $ f(\theta)$.
\begin{equation}
 \left[\frac{-\hbar^2}{2\mu}\nabla^2 + V_{eff}(r) \right]\psi( \vec r)=E \psi(\vec r)
 \label{eq_schrod}
\end{equation}
Total scattering amplitude $f(\theta)$ is expressed as the sum of Coulomb scattering amplitudes $f_C(\theta)$ and nuclear scattering amplitude $f_N(\theta)$ respectively thus,
\begin{equation}
f(\theta)=f_C(\theta)+f_N(\theta)
\end{equation}

The amplitudes $f_N(\theta)$ and $f_C(\theta)$ have expansions as follows.
\begin{equation}
 f_N(\theta)=\frac{1}{2ik}\sum_l(2l+1)e^{2i\sigma_l}\left(e^{2i\bar{\delta_l}-1} \right)P_l(cos\theta)
\label{eq_scattering2}
\end{equation}
\begin{equation}
 f_C(\theta)=\frac{1}{2ik}\sum_l(2l+1)\left(e^{2i\bar{\delta_l}-1} \right) P_l(cos\theta)
\label{eq_scattering2}
\end{equation}

Here k is the magnitude of wave vector $\overrightarrow{k}$, $\sigma_l$ is the Coulomb phase shift due to scattering and $\bar{\delta_l}$ is the nuclear phase shift. The ratio of the measured elastic scattering cross-section to Rutherford’s scattering cross-section is given by
\begin{equation}
\frac{d\sigma_{el}}{d\sigma_{Ruth}}=|{\frac{f(\theta)}{f_C(\theta)}}|^2
\label{eq_scattering3}
\end{equation}
For $ l ^{th}$ partial wave and its S-matrix $S_l $, the elastic scattering cross-section $\sigma_{el}$ and the reaction cross-section $\sigma_ r$ are given by Eq.12 and Eq.13 respectively as follows.
\begin{equation}
\sigma_{el}=\frac{\pi}{k^2}(2l+1)|1-S_l|^2
\label{eq_sigmael}
\end{equation}
\begin{equation}
\sigma_{r}=\frac{\pi}{k^2}(2l+1)T_l (E)=\frac{\pi}{k^2}(2l+1) (1-|S_l|^2)
\label{eq_sigmarl}
\end{equation}

Here, $T_l (E) = (1-|S_l|^2 $ . This is known as the transmission coefficient for the orbital angular momentum $ l $. The fusion cross section is given by, $\sigma_{fus}=\frac{\pi}{k^2}(2l+1)P_l ^F$ , where $P_l ^F$ is the fusion (absorption) probability. The wave function, in the case of fusion, is expected to be absorbed completely inside the barrier; hence the fusion probability can be assumed to be close to the probability that the incident current reaches the point of total absorption. 
Therefore, fusion probability, i.e.,$P_l ^F \sim T_l (E) = (1-|S_l|^2 $ . Thus, we have, $\sigma _r = \sigma _{fus}$. Based on the above theory and potential, the results of elastic scattering and fusion are discussed.
\begin{figure}[t]
\begin{center}
\includegraphics[width=10.0cm]{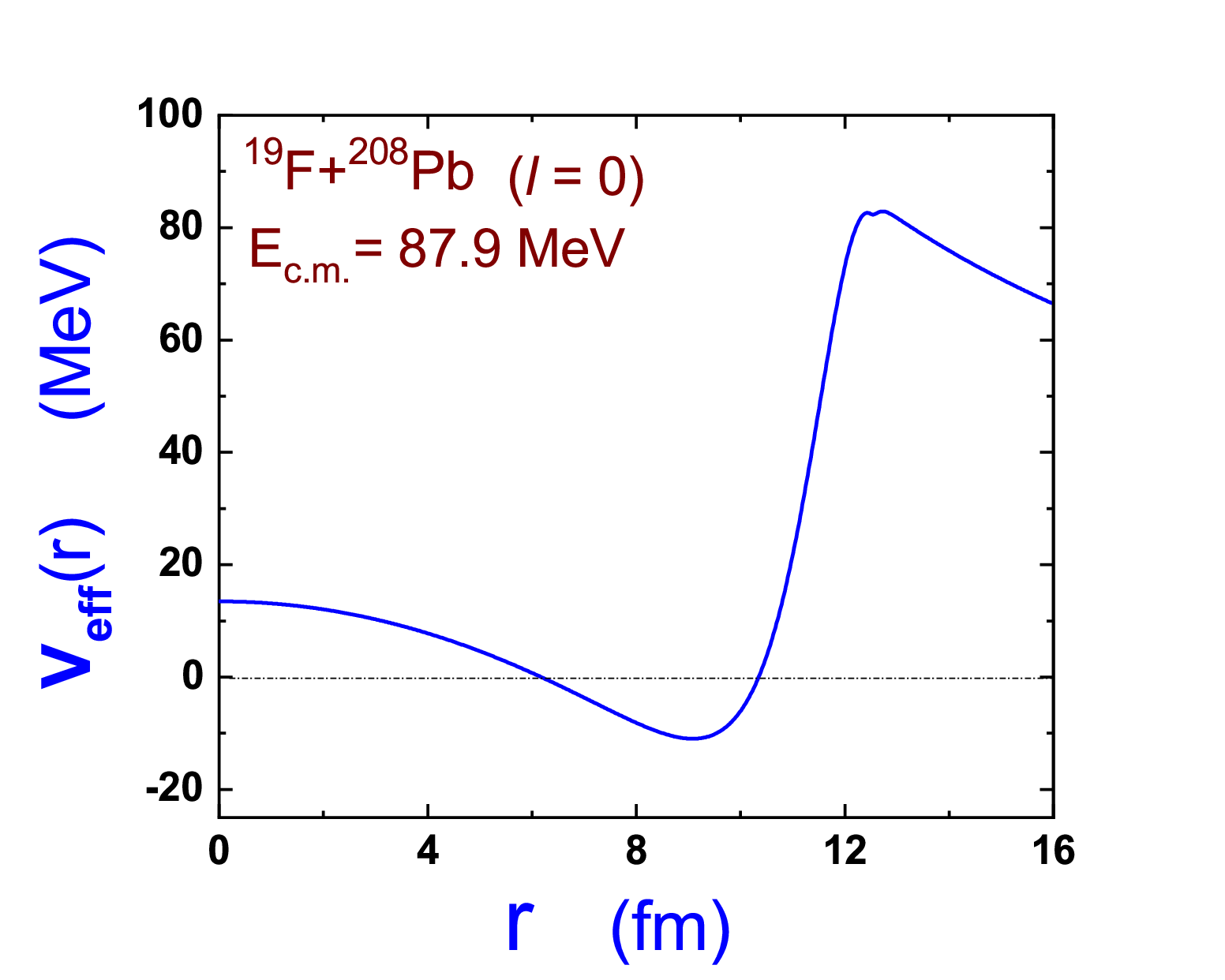}
\caption{%
The plot of real part of effective potential V$_{eff}$(r) for $l$= 0 of the system $^{19}F + ^{208}Pb$ at energy 87.9 MeV in the centre-of-mass frame. Values of parameters are $R_0$= 12.5 fm, $B_0$= 118 MeV, $B_1$= 7.0, $B_2$= 0.6 and $V_B$ = 2.6 MeV. The curve retains the neck structure at $R_0$.
}
\label{fig3}
\end{center}
\end{figure}
 \begin{figure}
     \centering     
         \includegraphics[width=0.5\textwidth]{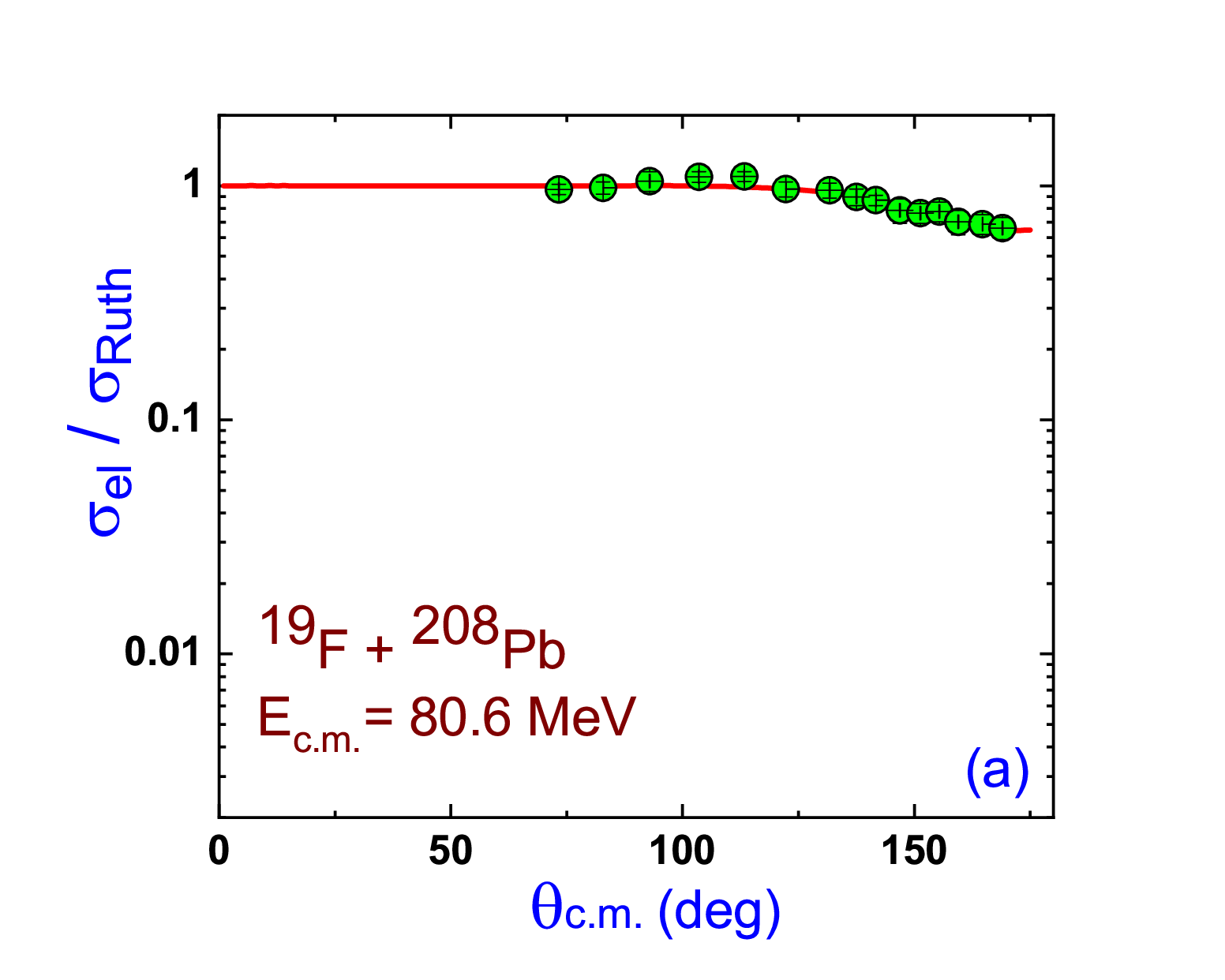}\includegraphics[width=0.5\textwidth]{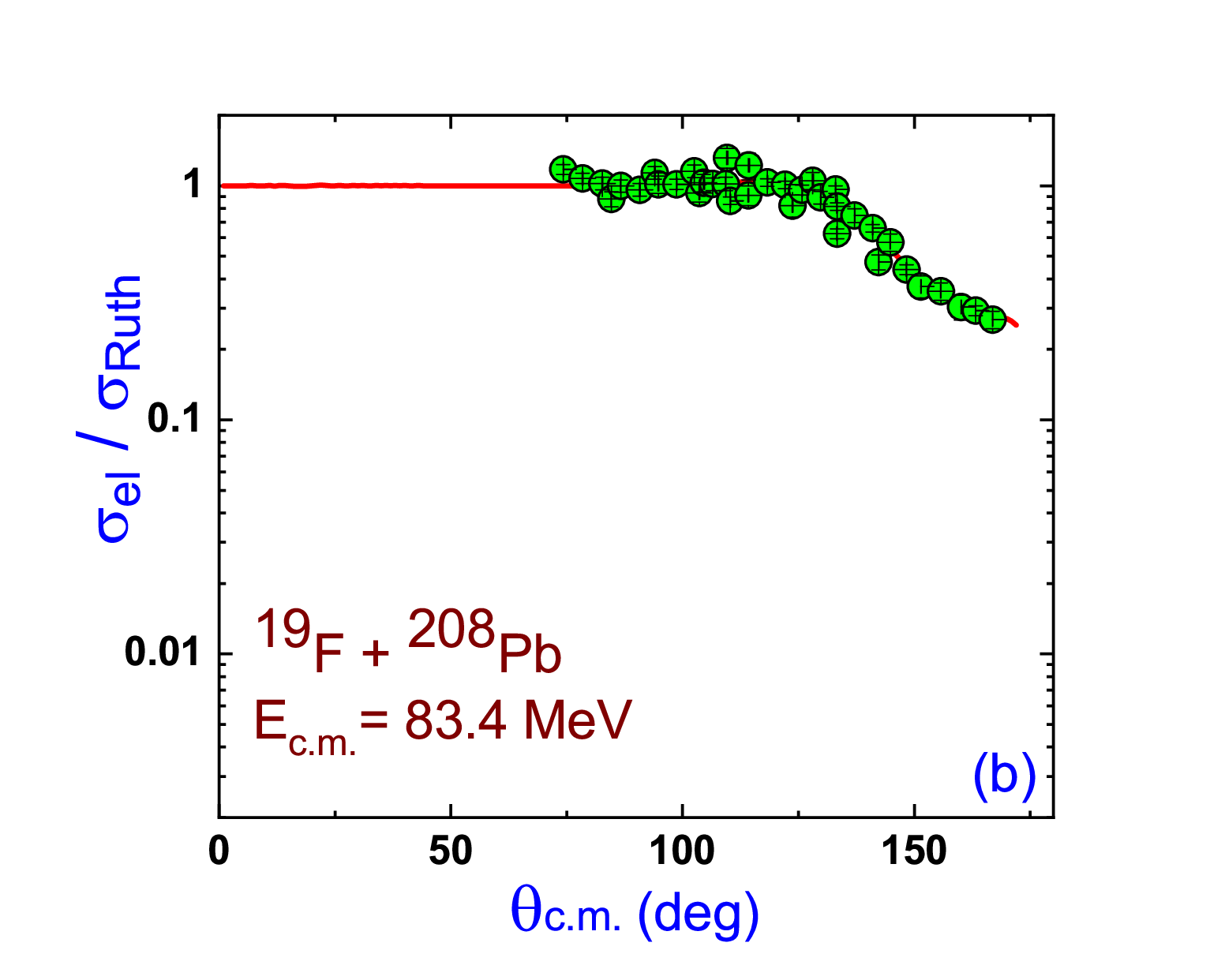}
          \includegraphics[width=0.5\textwidth]{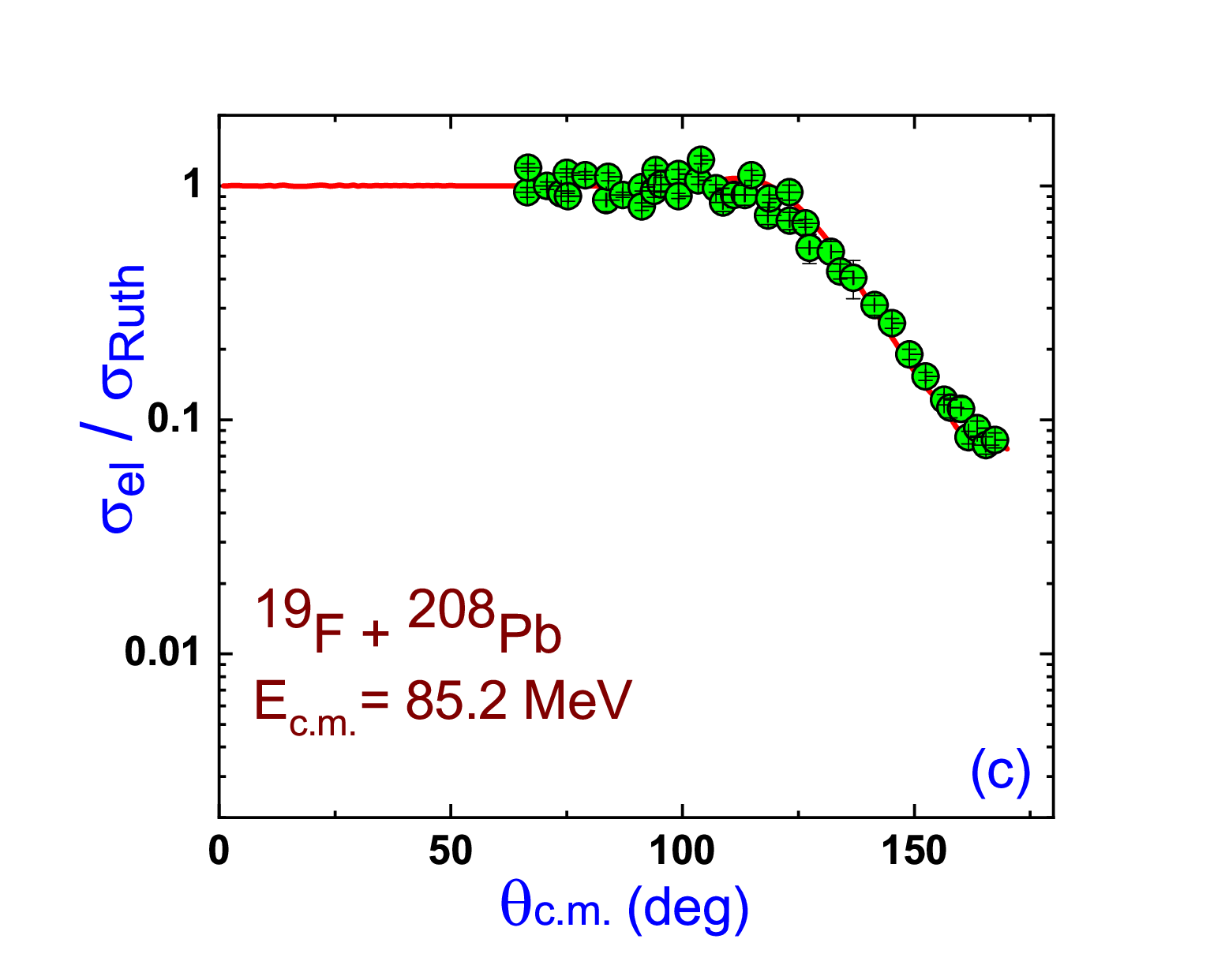}\includegraphics[width=0.5\textwidth]{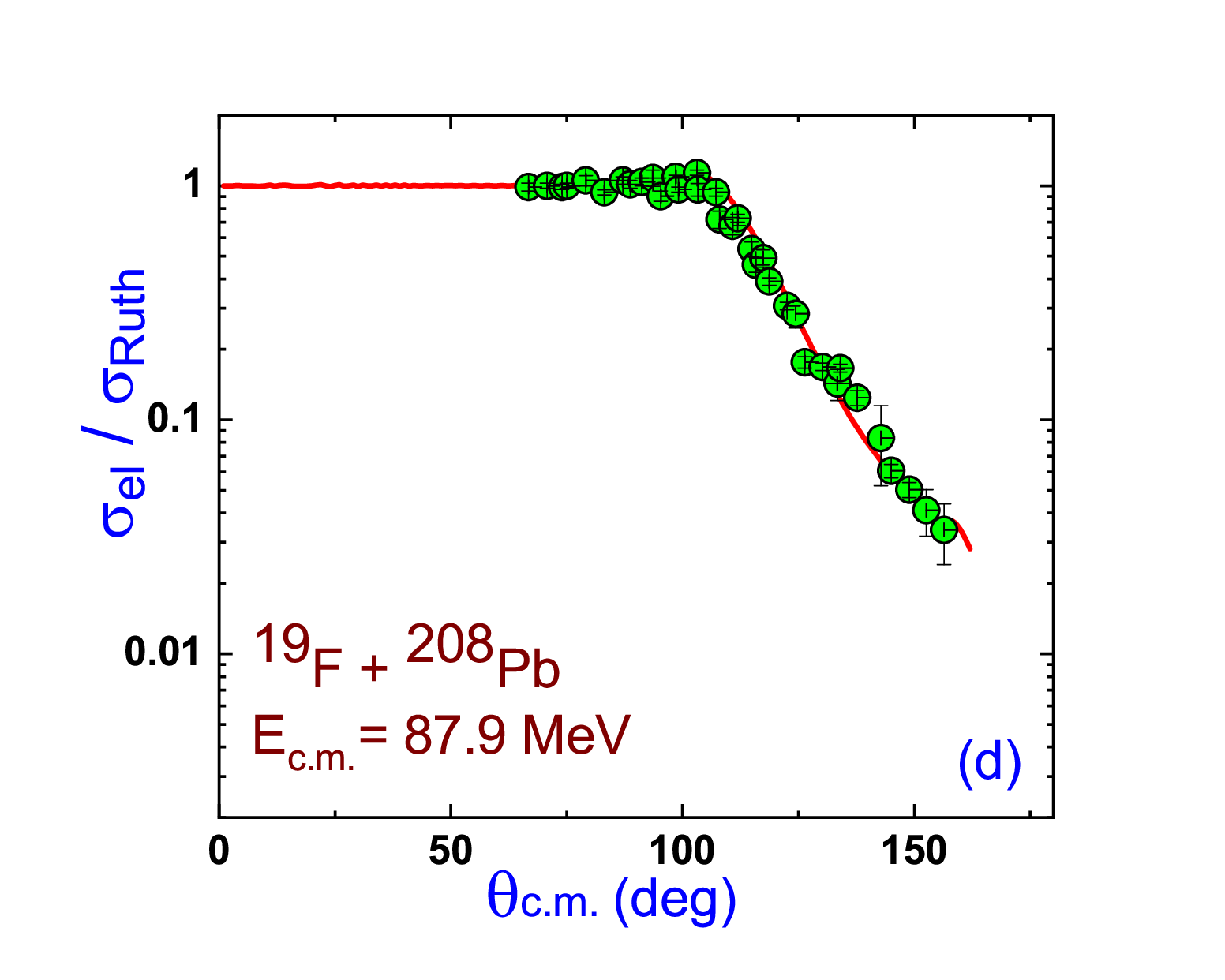}
          \includegraphics[width=0.5\textwidth]{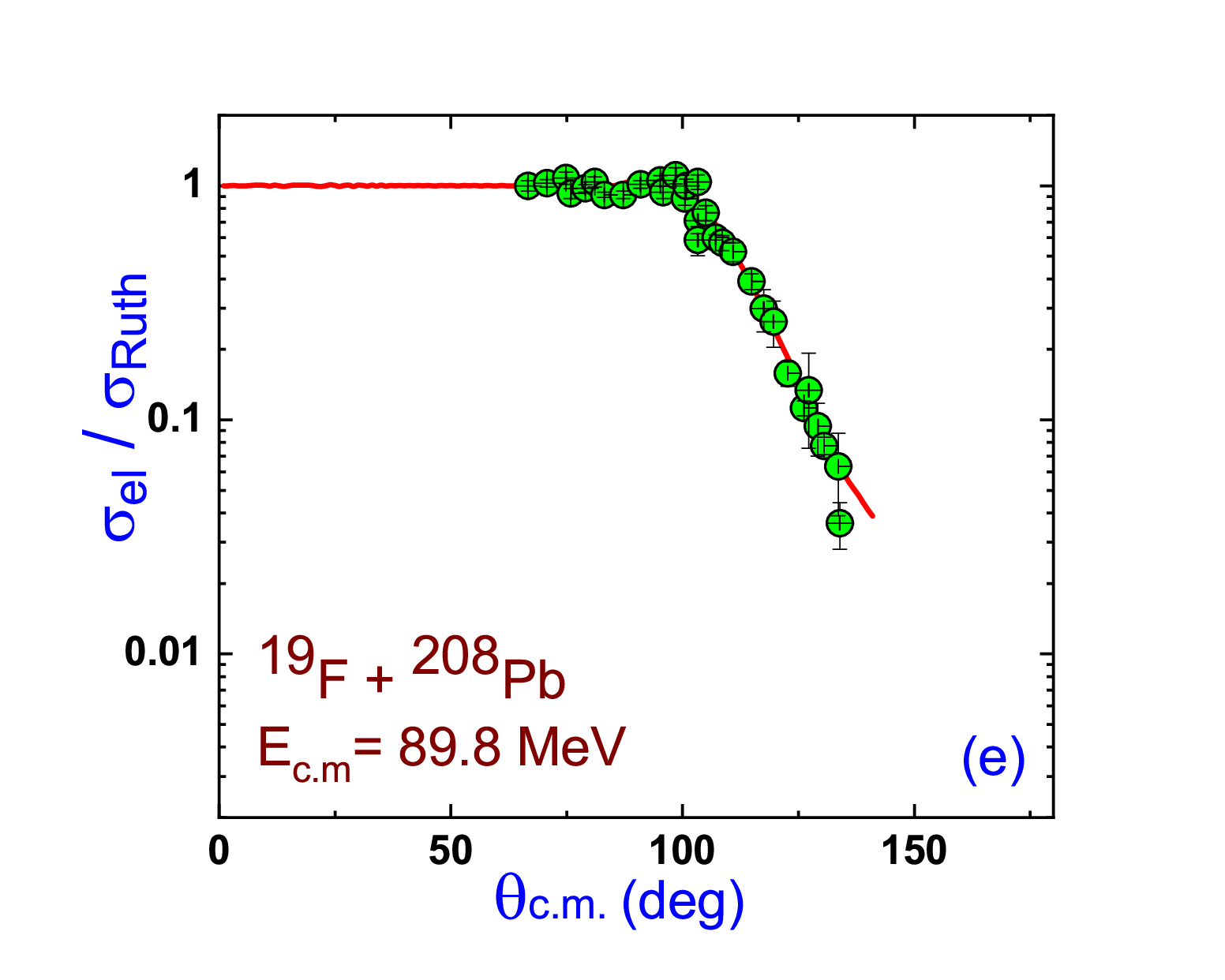}\includegraphics[width=0.5\textwidth]{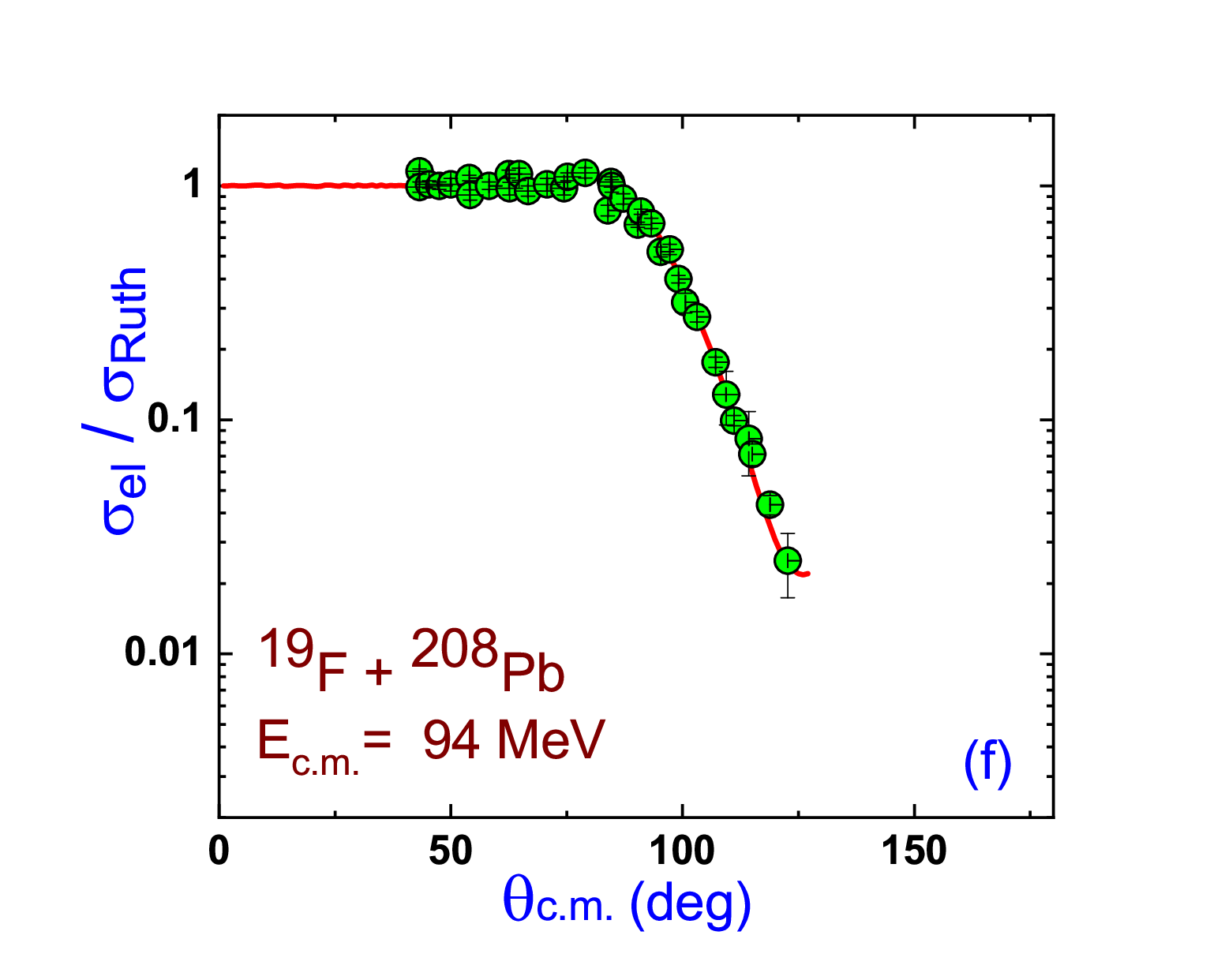}
\caption{Comparison of the calculated angular distribution of elastic scattering cross-sections with experimental data for six incident energies of the system $^{19}F +^{208}Pb$. The dark circles (green) represent experimental data and the solid curve (red) represents theoretical calculations. The experimental data for Ref.\cite{ref. 13} are downloaded from the site http: //nrv.jinr.ru. The values of the six independent parameters are found $R_0$= 12.5 fm, $R_{0W}$ = 12.8 fm, $B_2$= 0.6, $B_0$= 118 MeV, $W_1$=1.4 and $W_0$=1.5 MeV for all incident energies.}
\label{fig4}
\end{figure}
\section{RESULTS}
While analyzing angular distribution cross-sections of the elastic scattering $^{19}F+^{208}Pb$, C. J. Lin et al. \cite{ref.13} have used Woods-Saxon based optical potential, in which imaginary potential depths are approximately 29 - 60$\%$ or more of the real part to get the best fit of data for energy span 87.9 - 94.0 MeV. The minimum value of the real part differs by 21$\%$ from the maximum value, and the difference is more than 96$\%$ in the case of imaginary parts for the same range of incident energy. Scattered values may raise uncertainty to conclude threshold anomaly and higher imaginary values concerning real parts may suppress resonance states of the system generated by the effective potential. We use a substantially small imaginary part to real part ratio to measure angular distribution cross-sections at different incident energies. The success story of the potential generated by using Ginocchio potential \cite{ref.21, ref.22, ref.23} catalyzes us to explain the experimental data in this optical model analysis. The variations in real and imaginary parts near the Coulomb barriers are studied to realize threshold anomaly. The results are presented in the following sub-sections.
 
\subsection{Analysis of scattering cross-sections}
The laboratory energies for elastic scattering of collimated $^{19}F$ beam by $^{208}Pb$ target are taken at 88, 91, 93, 96, 98, and 102 MeV, which are equivalent to 80.6, 83.4, 85.2, 87.9, 89.8, and 94.0 MeV respectively in the center of mass frame. The system $^{19}F + ^{208}Pb$ has the Coulomb barrier at about $E_{c.m.}$= 84 MeV. Energy-dependent parameters (both real and imaginary) of the optical potential are presented in Table-I for the best fit. The calculated angular distribution of elastic cross-sections is compared with experimental values in Fig.4 for the given range of incident energies. The experimental values digitized with GSYS-2.4 are obtained from the source http ://nrv.jinr.ru.

\begin{table}
\caption{Energy-dependent parameters of the Optical Potential}
\renewcommand{\arraystretch}{1.8}
 \centering
 \small\addtolength{\tabcolsep}{8pt}
 \begin{tabular}{ |c|c|c|c|c|c| }
  \hline
    $E_{CM}$ & $V_B$ & $W_2$ & $V_{BW}$ & $B_1$ \\
  (MeV)& (MeV) &  & (MeV) & \\
  \hline
  80.6 & 2.2 & 2.5 & 0.06 & 2.5 \\
  \hline
  83.4 & 2.4 & 2.2 & 0.1 & 6.0 \\
  \hline
  85.2 & 2.5 & 1.2 & 0.3 & 6.0\\
  \hline
  87.9 & 2.6 & 0.001 & 0.6 & 7.0 \\
  \hline
  89.8 & 2.3 & 2.5 & 0.7 & 2.0\\
  \hline  
  94.0 & 2.2 & 3.0 & 1.2 & 1.0 \\
  \hline
 \end{tabular}
 \label{table_paramet_al} 
\end{table}
Six number of parameters show energy-independence. The radial distance in the surface region $R_0$ is kept constant with energy. The values of the independent parameters are $R_0$ = 12.5 fm, $R_{0W}$ = 12.8 fm, $B_2$ = 0.6, $B_0$= 118 MeV, $W_0$=1.5 MeV and $W_1$=1.4 while matching theoretical results with experimental outcomes for the entire range of energies. Four parameters, namely, $B_1$, $V_B$, $W_2$, and $V_{BW}$ mentioned in Table-I vary with collision energies. We keep the Coulomb radius parameter $r_C$ at 1.33 fm following the literature referred \cite{ref.13}. The theoretical data (solid line curve in red colour) are compared with the experimental data (dark circles filled with green colour) obtained from Ref. \cite{ref.13}. Theoretical calculations fairly agree with the experimental values. \\

It is worth mentioning that the best fit takes a smaller imaginary part in comparison to the corresponding real part of the potential which ensures less suppression of resonance states generated by the effective potential. The real parts of the potential are taken the same, i.e., 118.0 MeV for all the six incident energies. The imaginary parts are also kept small and the same, i.e., 1.5 MeV in comparison to their counterpart's real potential. Thus, the ratios of imaginary parts to real parts remain the same, i.e., 0.0127 for all colliding energies. Such a small imaginary-to-real ratio will help not to substantially destroy the resonance states generated by the volume part of the effective potential. The need for a small imaginary part in an optical potential is explained with supporting graphs in sub-section 3.4.

 \begin{figure}
     \centering
         \centering
         \includegraphics[width=0.7\textwidth]{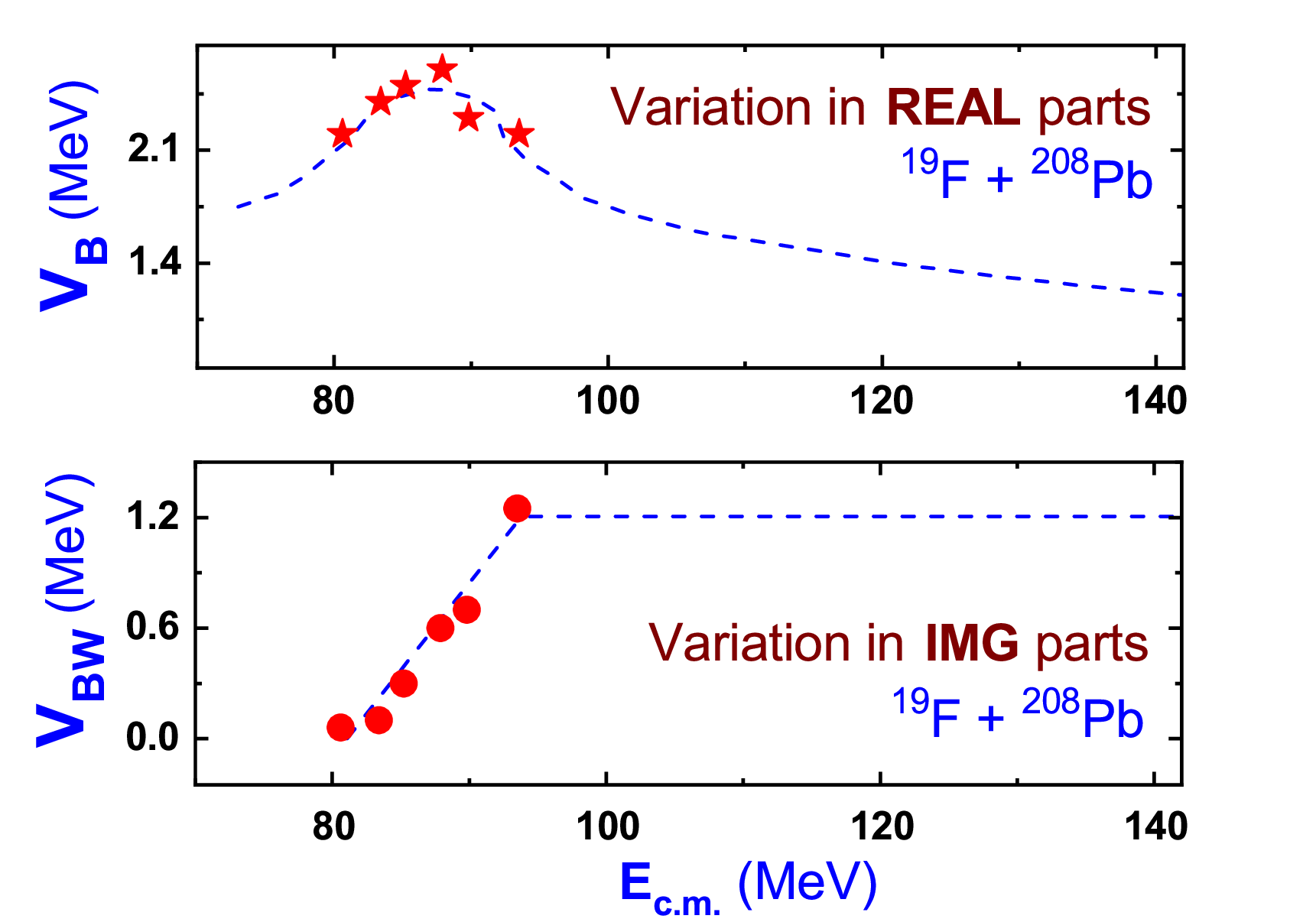} 
    \caption{Variation of real part ($V_B$) and imaginary part ($V_{BW}$) of the optical potential around the Coulomb barrier. The blue dash-line with red star-symbols represents the variation of real parts with incident energies, whereas, the blue dash-line with dark red circles represents simultaneous variation of imaginary parts. The variation shows the presence of TA in the vicinity of the Coulomb barrier.}
    \label{fig5}
\end{figure}
\subsection{Phenomenon of threshold anomaly}
While reproducing the experimental results using theoretical calculations of the potential, we find variations in real and imaginary parts of the potential near the Coulomb barrier. The variations are described in Fig.5. When we proceed from the lower energy of the collision, the real part first increases and then decreases in the vicinity of the Coulomb barrier, and ultimately saturates around 2.2 MeV at higher energies away from the barrier. On the other hand, the imaginary part remains almost constant at 1.2 MeV at higher energies but decreases in the vicinity of the Coulomb barrier as we move from higher to lower values of incident energy. The variation in the real part follows a bell-shaped dash line (upper plot), whereas, the variation in the corresponding imaginary part follows an L-shaped dash line (lower plot). The bell-shaped and L-shaped curves are obtained by the dispersion relations in Ref.\cite{ref.13}, which describes threshold anomaly. Thus, theoretical calculations fairly agree with the TA phenomenon described in the reference.

\subsection{Analysis of fusion cross-sections}

Along with the angular distribution, the present analysis is extended with the potential to check the fusion cross-sections. The experimental data of the fusion cross-section ($\sigma_{fus}$) with $^{19}F+^{208}Pb$ system are available by different groups from independent experiments. B. B. Back et al. \cite{ref.25} measured the fission fragment angular distributions in 1985 by using Argonne Superconducting LINAC in the energy range 100.8 to 174.0 MeV in the center-of-mass frame. The fission cross sections and angular distributions were measured in 1990 by Zhang Huanqiao et al. \cite{ref.27} with HI-13 tandem Van de Graff accelerator at bombarding energies from 75.51 to 137.5 MeV. The fusion-fission cross-sections were also measured in 1998 by K. E. Rehm et al. \cite{ref.26}  by using ATLAS superconducting linear accelerator for the reaction at energies from 77.68 to 99.6 MeV. 1n 1999, D. J. Hinde et al. \cite{ref.24} were able to measure fission fragment cross sections and angular anisotropies to high accuracy for the reaction by using 14UD tandem electrostatic accelerator and LINAC in the energy range 76.0 to 144.7 MeV. The experimentally measured fusion cross-section values digitized with GSYS-2.4 are downloaded from the website http://nrv.jinr.ru. \\

To explain the data of the fusion cross-section, we use the same parameters except for $R_{0W}$ and $W_2$ of the optical potential obtained while matching theoretical calculations with the corresponding angular distributions in the scatterings of the system at the incident energy of 94 MeV in the center-of-mass frame. The value of $R_{0W}$ has been altered from 12.8 MeV to 8.2 MeV and that of $W_2$ from 3.0 to 0.8 to explain fusion data. Thus, the set of parameters to reproduce fusion cross-sections is $R_0$=12.5 fm, $R_{0W}$ =8.2 fm, $B_1$=1.0, $B_2$=0.6, $V_B$=2.2 MeV, $V_{BW}$=1.2 MeV, $B_0$= 118 MeV, $W_0$= 1.5 MeV, $W_1$=1.4 and $W_2$= 0.8. The set of parameters fairly explains the experimental values of four independent experiments performed for different energy ranges. The results are shown in Fig.6 with fusion cross-sections (expressed in mb) as the function of the center of mass energy (taken in MeV). It is challenging to find a unique potential that can address both of these phenomena simultaneously. $R_{0W}$ in case of fusion is considered to be 8.2 fm, which is less than the Coulomb barrier position ($R_B$ = 12.8 fm) keeping the imaginary $W_0$ unchanged at 1.5 MeV to observe the structure phenomena. The value of $R_{0W}$ 
($ < R_B$) confirms that fusion takes place only after the barrier has been fully penetrated. The other parameter $W_2$ was reduced from 3.0 to 0.8 to observe the fusion phenomena. Finally, we get the required fusion cross-section which is plotted in Fig.6. The structure effect is also visible here because $^{208}Pb$ is a double magic and shell closure nucleus and $^{19}F$ is relatively stable projectile. The binding energy per nucleon value of $^{19}F$ is higher compared to other isotopes such as $^{21}F$ and $^{23}F$. We do not find hindrances below the Coulomb barrier because of the shell closure of both the projectile and target. As we move toward the neutron drip-line we see the hindrance phenomena as described in Fig.7.

\begin{figure}
\begin{center}
\includegraphics[width=10.0 cm]{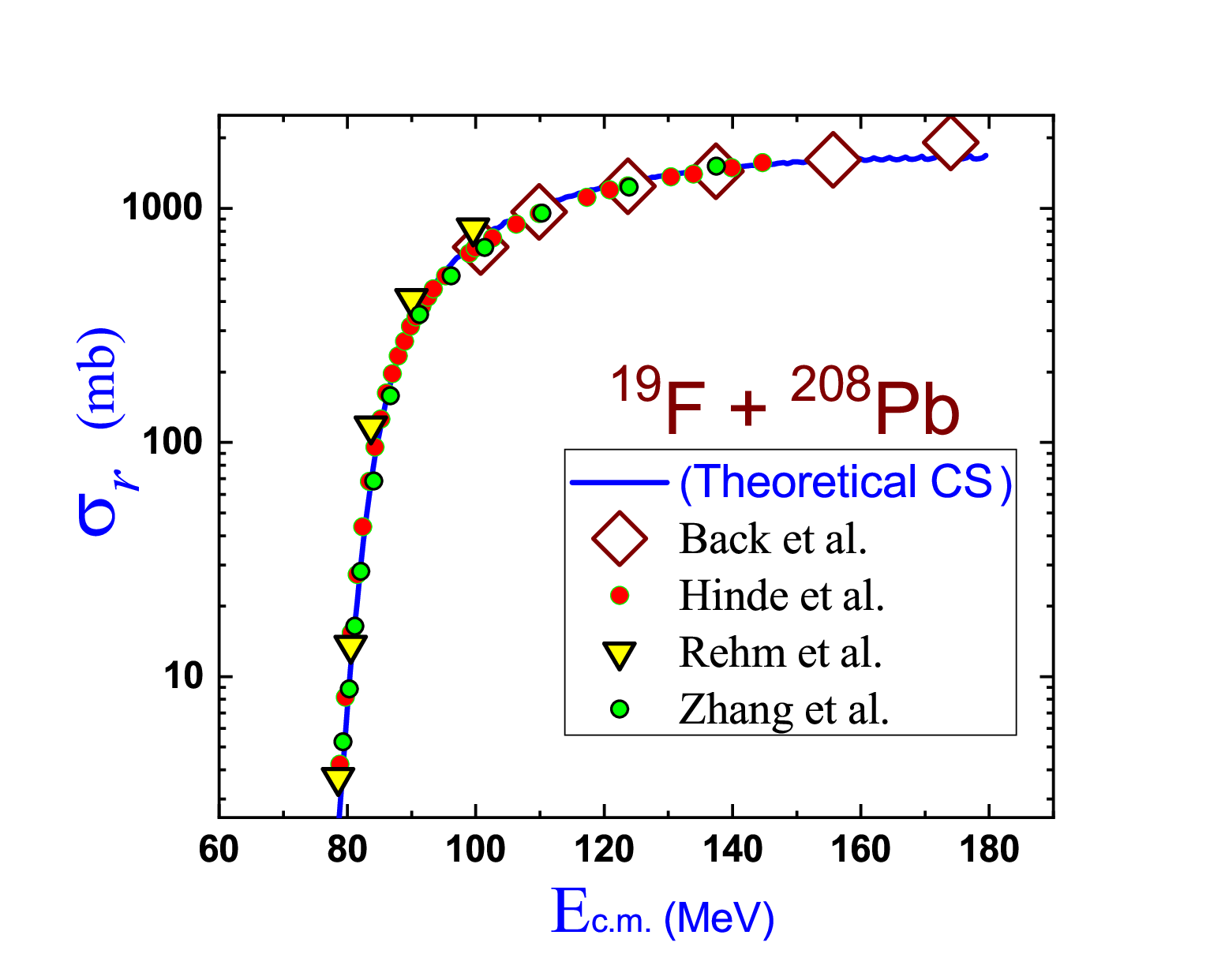}
\caption{%
Plot of fusion cross-sections as a function of colliding energy in centre-of-mass frame. The solid curve (blue) represents theoretical values of fusion cross-sections. The experimental values obtained from Back et al. are represented by rhombus, Hinde et al. by red circles, Rehm et al. by triangles and Zhang et al. by green circles. The experimental data are downloaded from the site http://nrv.jinr.ru. Set of parameters taken are $R_0$=12.5 fm, $R_{0W}$ =8.2 fm, $B_1$=1.0, $B_2$=0.6, $V_B$=2.2 MeV, $V_{BW}$=1.2 MeV, $B_0$= 118 MeV, $W_0$= 1.5 MeV, $W_1$=1.4 and $W_2$= 0.8. }
\label{fig6}
\end{center}
\end{figure}
\begin{figure}
\begin{center}
\includegraphics[width=10.0cm]{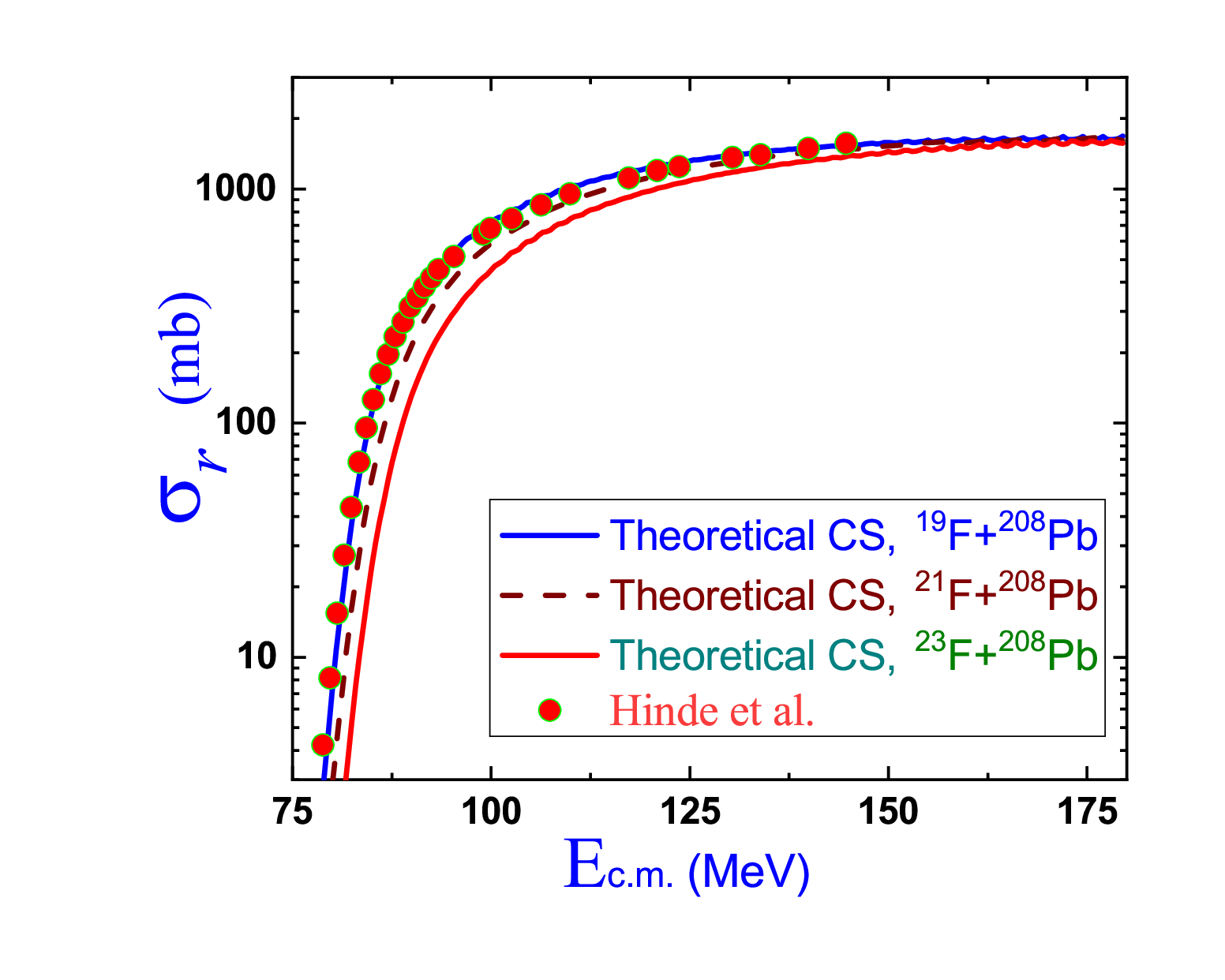}
\caption{%
Plots of fusion cross-sections with different isotopes of Fluorine projectile. The figure shows the structure effect due presence of double magic and shell closure nucleus.
 }
\label{fig7}
\end{center}
\end{figure}
It is commonly known that a projectile's binding energy has a significant impact on how easily it fragments into smaller parts, which has an impact on the fusion of weakly bound projectiles. It is difficult to calculate the dynamic polarization potential theoretically, thus up to now, TA has only been studied experimentally. Most of the examined systems are spherical or very close to spherical shape. The implications of nuclear structure on well-deformed systems have received very little attention in research. C. J. Lin et al. \cite{ref.13} used the system $^{19}F+^{208}Pb$ to investigate the role of deformed nuclei in fusion and TA responses. The projectile nucleus $^{19}F$ possesses \cite{ref.36} quite large static deformations ($\beta_2$ = 0.44 and $\beta_4$ = 0.14). The results of the fusion reaction \cite{ref.13} for the system have been carefully examined. We analyze the cross-sections of fusion reactions for the fluorine isotopes with the double shell closure of $^{208}Pb$. The nucleus of a nearby projectile with double shell closure is $^{16}O$. In the case of $^{19}F$, one half-filled valence proton and two valence neutrons are present, and the same is evident for $^{21}F$ and $^{23}F$ nuclei. As we approach the drip-line nuclei, the two-neutron separation energy ($S_{2n}$) of fluorine continues to decrease. The magnitudes of $S_{2n}$ for the isotopes $^{19}F$, $^{21}F$ and $^{23}F$ are 19.582 MeV, 14.742 MeV, and 12.81 MeV respectively \cite{ref.37}. Moreover, the corresponding binding energies per particle (B.E./A) are 7.779 MeV, 7.738 MeV, and 7.622 MeV respectively. Again, Z=82 and N=126 in the target nucleus are double magic numbers. The isotopes of Fluorine (F) could split into $^{16}O$ and neutron clusters due to the valence neutrons' low binding energy. 

Again, it is observed that the presence of magic shells in the entrance channel increases the probability of fusion \cite{ref.38}, because magic nuclei are difficult to excite, which lowers energy dissipation and facilitates the creation of a more compact di-nuclear system. So, when we go away from the shell closure, the 2n, 4n, and 6n evaporation residue cross-sections may indeed be enhanced close to the Coulomb barrier, which might account for the hindrance of the fusion cross-section shown in Fig.7. Nevertheless, it is beyond the scope of our approach to demonstrate the entire reaction cross-section, including quasi-fission and evaporation residues. We have just displayed the fusion cross-section here. Because of quasi-fission, the fusion probability could well be significantly suppressed below the Coulomb barrier, which might be the cause of the fusion hindrance \cite{ref.39}. Hence, the fusion of weakly bound projectiles is affected by the breakup channel coupling. Fig.7 illustrates how the breakup channel coupling affects $^{19-23}F+^{208}Pb$ systems depending on the valence nucleon separation energy. As per expectation, the breakup cross-section grows as the binding and separation energies fall, making the breakup of a weakly bound nucleus more feasible. As the structure of interacting nuclei influences the mechanism of the fusion and other processes leading to the absorption of particles from the elastic channel then it is natural to expect that the Optical Model parameters should vary from one system of colliding nuclei to another one. The parameter $R_0$ is altered. So the potential barrier and hence the fusion cross section increase. We demonstrate the effects of only charge and mass on the fusion phenomenon. Nuclear shell structures, deformation, and orientation are additional factors that influence the fusion processes in addition to charge and mass \cite{ref.28,ref.29,ref.30}.

\subsection{Need for a small imaginary part in the optical potential}

The scattering process in a nuclear collision is sensitive to the nature of the potential on the surface region. On the contrary, the fusion process is an interior activity. It is quite difficult to find a unique nuclear potential that can take care of both phenomena. \\

It is a common assumption that fusion takes place only after the barrier has been fully penetrated \cite{ref.40, ref.41}. Based on this concept that the fusion of two nuclei occurs in the region interior to the radial position ($R_B$) of the Coulomb barrier, the region $0 < r < R_B$ is expected to account for the experimental data of fusion cross section as the total reaction cross section includes the cross section for different reaction channels of which the fusion channel is predominant in the low-energy collision activities. The values of fusion radius and the Coulomb radius used in the heavy ion collision system $^{19}F+^{208}Pb$ agree with the fact that fusion is an interior phenomenon, whereas the surface phenomenon is attributed to scattering and other peripheral, less absorptive direct reaction processes. \\

In this study, we identify two crucial aspects of our potential: (i) the real component with a larger magnitude, and (ii) the imaginary part with a smaller value. Thus, in contrast to light ion systems, this potential has a less absorptive character. Because of its less absorptive nature, standing waves are formed in the nuclear well, which allows shape resonances to survive in the collision process. As a result, these resonances produce the oscillatory structures in the fusion (total reaction) cross-section, $\sigma_{fus}$ ($\sigma_r$) as a function of colliding energy $E_{c.m.}$ Although the resonances exist, it is very difficult to detect the resonances experimentally through direct observations \cite{ref.42}. \\
\begin{figure}
\begin{center}
\includegraphics[width=10.0cm]{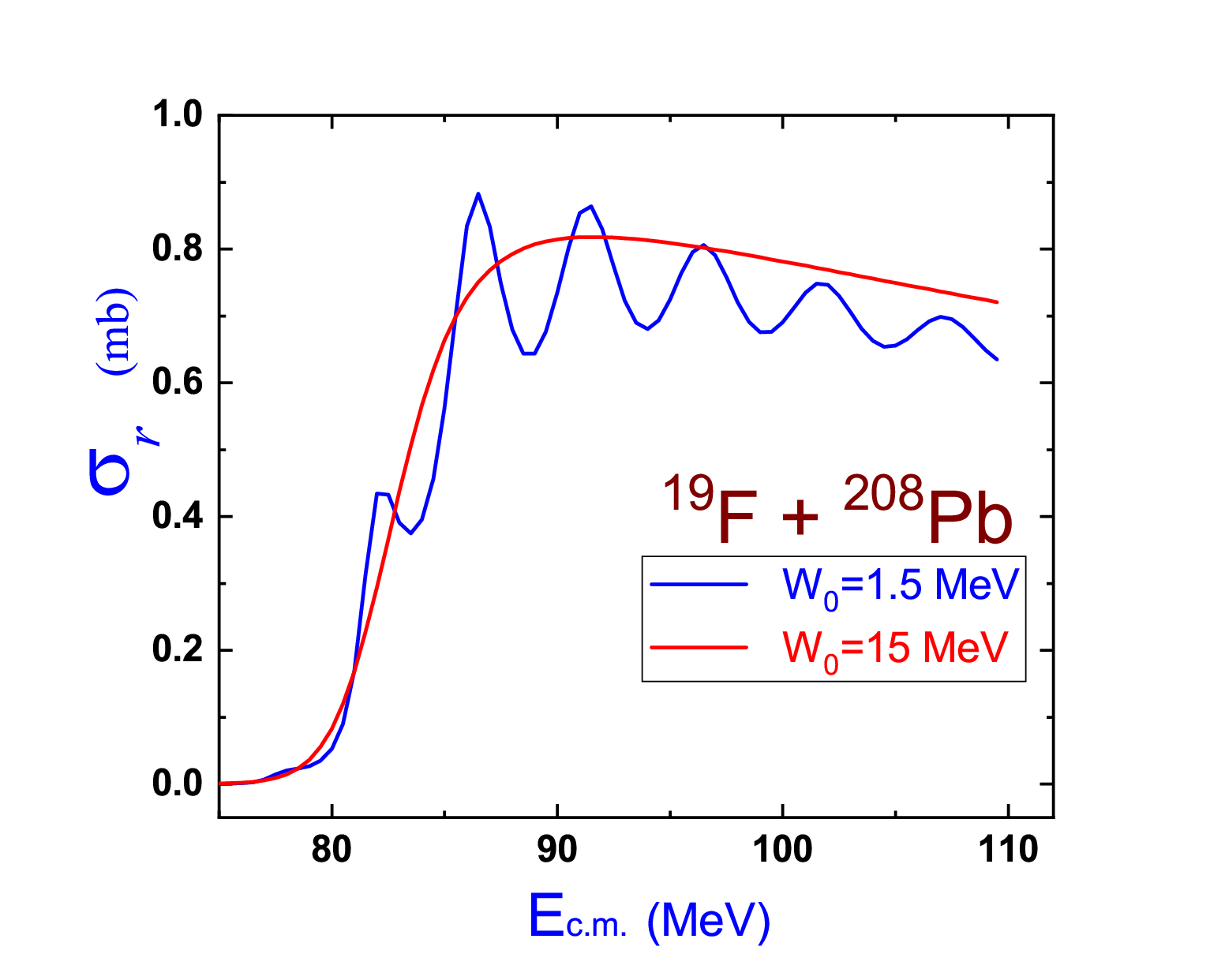}
\caption{%
Comparison of reaction cross section curve at $W_0$=1.5 MeV with reaction cross section curve at $W_0$=15 MeV keeping other parameters unaltered. Resonances are shown in the case of a lower $W_0$ value (blue curve).
 }
\label{fig8}
\end{center}
\end{figure}
\begin{figure}
\begin{center}
\includegraphics[width=10.0cm]{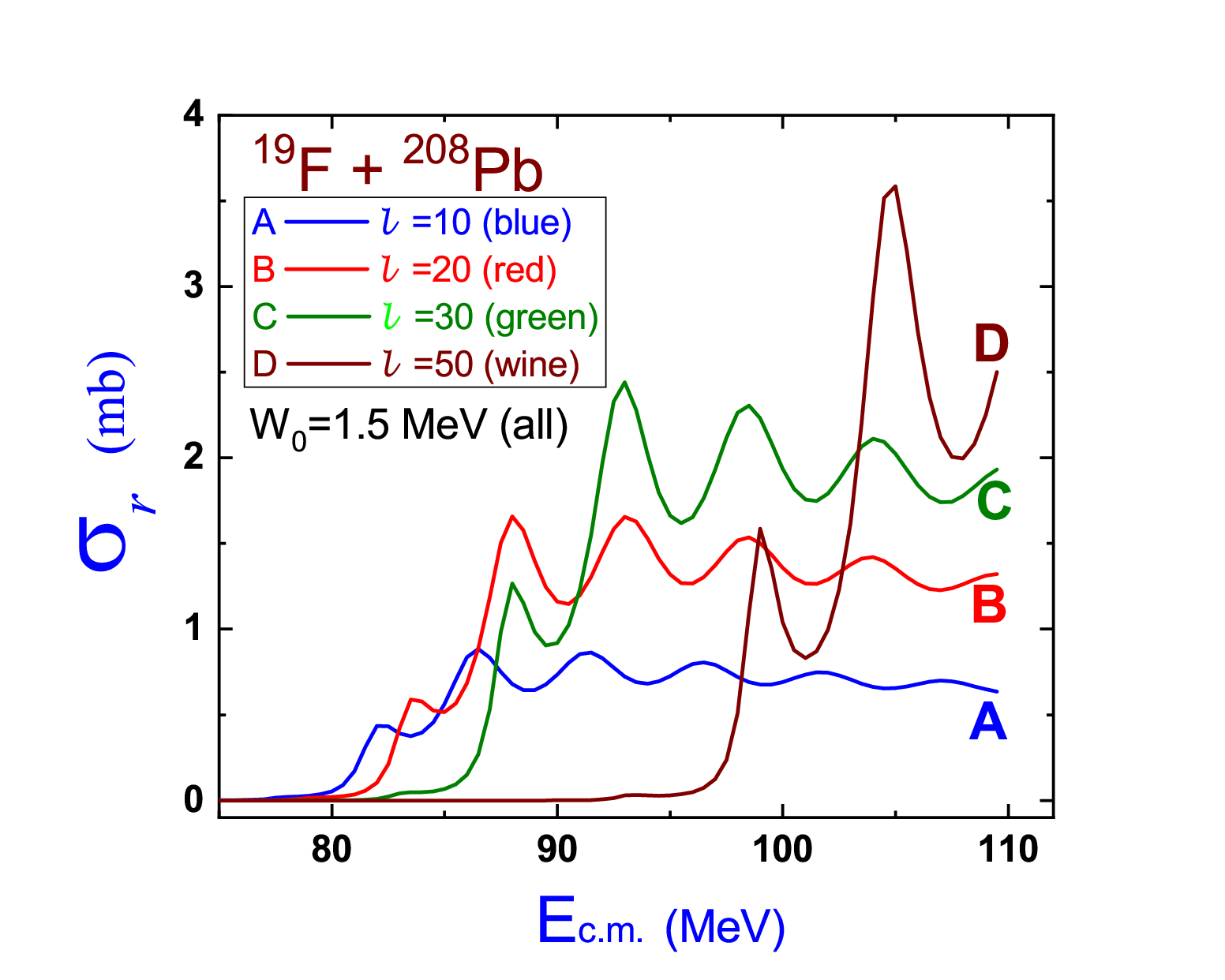}
\caption{%
Reaction cross-section curves for the system $^{19}F + ^{208}Pb$ against incident energy. Resonances are shown in case of lower $W_0$ value, i.e., $W_0$=1.5 MeV keeping other parameters unaltered except the values of $ l $. The amplitude of oscillation increases with the increase of $l$-values of partial waves.
 }
\label{fig9}
\end{center}
\end{figure}
In the potential scattering theory, these resonances are manifested clearly as maxima in the results of reaction cross-section ($\sigma_r$) at the respective resonance energies \cite{ref.43, ref.44}. This small value of the imaginary part further indicates that the fusion only occurs when the barrier has been completely penetrated \cite{ref.41, ref.45, ref.46}. Due to the potential's smaller absorption capacity, standing waves in the nuclear well might occur, which would allow shape resonance states (which have not been experimentally detected) \cite{ref.42} to survive the collision process. As a result, these resonances take on the role of being the cause of the oscillatory structure in the barrier distribution, $D(E_{c.m.})$ findings as a function of $E_{c.m.}$ \cite{ref.45, ref.46}. When the potential is made more absorbed by considering a bigger imaginary part $W_0$, the width of the resonance caused by the real part of the potential widens. Consequently, larger width leads to the extinction of the corresponding resonance in the collision process. In this study, we have considered a deep real potential associated with a relatively weak imaginary strength $W_0$. As explained above, the resonances are visible in the form of peaks in the partial wave trajectories for a smaller value of $W_0$, but the oscillation in $\sigma_r$ vanishes for a larger value of $W_0$, which is shown explicitly by taking $W_0$=1.5 MeV and 15 MeV in Fig.8. The cumulative effect of all these resonances is primarily responsible for the oscillation in reaction cross-section ($\sigma_r$). The fusion radius is found to be more than the Coulomb radius when a larger value of $W_0$ is considered in the case of heavy ion collisions. \\

The amplitudes of oscillation increase with increasing $ l $-values for a particular low imaginary potential. This is verified by changing the variation of resonance structures for different $l$’s with a particularly low value of $W_0$ as shown in Fig.9. The plots in Fig.9 explain how the amplitudes of oscillation increase with increasing $ l $-values for a particular low imaginary potential. Here the imaginary depth is kept low at $W_0$=1.5 MeV and the oscillations with increasing amplitudes are shown for different values of $ l $, i.e., $ l $=10, 20, 30, 50.

\section{CONCLUSIONS}
We use the optical potential taken into consideration in the paradigm of Ginocchio potential to explain the angular distributions of elastic scattering of the system $^{19}F+^{208}Pb$ for the center of mass energies $E_{c.m.}$= 80.6, 83.4, 85.2, 87.9, 89.8, and 94.0 MeV. At a nucleus' surface, the potential has a particular deformation effect. We calculate the fusion cross-sections for the same system and compare these values with various independent findings from four distinct experiments carried out by the researchers in Ref. \cite{ref.24, ref.25, ref.26, ref.27}. A theoretical calculation provides a good explanation for the data showing threshold anomaly close to the system's Coulomb barrier. To ensure that resonance states are not too suppressed, the imaginary components of the potential employed in the current study are kept relatively modest in comparison to the real parts. Due to the shell closure of both interacting nuclei, no hindrance phenomenon is seen in the system $^{19}F+^{208}Pb$. \\

As shown in Fig.7, a weakly bound nucleus significantly impacts the fusion due to the increased possibility of dissociation. So, it may be argued that although the Coulomb repulsion is stronger, the existence of neutron shell closure and breakup probability in the entrance channel favours fusion hindrance just below the Coulomb barrier. More studies are needed to understand the dynamics in the sub-barrier area and to identify additional influencing elements that may further favour or hinder the likelihood of fusion, such as deformation of the colliding partners, projectile direction upon striking the target, isospin asymmetry of the colliding partners and shell energy \cite{ref.47} etc. So it has not only the kinematical origin governed by the atomic mass and charge number, but it could be due to shell structures, deformation and shell energy \cite{ref.31, ref.32, ref.47}. The adaptability of the potential with less energy-dependent parameters encourages further analysis of more pertinent systems.



\begin{thebibliography}{99}

\bibitem{ref.1} M.A. Nagarajan, C. C. Mahaux, and G. R. Satchler, Phys. Rev. Lett. {\bf 54}, 1136 (1985).
\bibitem{ref.2} J. Diaz, J.L. Ferrero, J.A. Ruiz et al., Nucl. Phys. A {\bf 494}, 311 (1989).
\bibitem{ref.3} B. R. Fulton, D.W. Banes 1, J.S. Lilley et al., Phys. Lett. B {\bf 162}, 55 (1985).
\bibitem{ref.4} M. E. Brandan, J. R. Alfaro, A. Menchaca-Rocha et al., Phys. Rev. C {\bf 48} (1993) 1147.
\bibitem{ref.5} M. J. Smithson, J.S. Lilley, M.A. Nagarajan et al., Nucl. Phys. A {\bf 517} (1990) 193-204.
\bibitem{ref.6} A. Baeza,  B. Bilwes, R. Bilwes et al., Nucl. Phys. A {\bf 419} (1984) 412.
\bibitem{ref.7} I. J. Thompson, M.A. Nagarajan, J.S. Lilley et al., Nucl. Phys. A {\bf 505} (1989) 84-102.
\bibitem{ref.8} J.S. Lilley, B.R. Fulton, M.A. Nagarajan et al., Phys. Lett. B {\bf 151} (1985) 181-184.
\bibitem{ref.9} A.M. Stefanini, D. Bonamini, A. Tivelli et al., Phys. Rev. Lett. {\bf 59} (1987) 2852.
\bibitem{ref.10} D. Abriola, D. DiGregorio, J. E. Testoni et al., Phys. Rev. C {\bf 39} (1989) 546.
\bibitem{ref.11} G. R. Satchler, Physics Reports, North-Holland, {\bf 199}, 3 (1991) 147-190.
\bibitem{ref.12} F. W. Byron and R. W. Fuller, Math. of Class. and Quant. Physics, (1992) 340.
\bibitem{ref.13} C. J. Lin, J. C. Xu, H. Q. Zhang et al., Phys. Rev. C {\bf 63}, 064606 (2001).
\bibitem{ref.14} H. Leucker, K. Becker, K. Blatt et al., Phys. Lett. B {\bf 233}, 277 (1989).
\bibitem{ref.15} D. R. Tilley, H.R. Weller, C.M. Cheves et al., Nucl. Phys. A {\bf 595}, 1-170 (1995).
\bibitem{ref.16} Amit Kumar, R. Tripathi, S. Sodaye et al., Euro. Phys. Journal, A {\bf 49}, 3 (2013).
\bibitem{ref.17} U. C. Voos, W. Von Oertzen, R. Bock et al., Nucl. Phys. A {\bf 135}, 207-224 (1969).  
\bibitem{ref.18} U. C. Schlotthauer-Voos, H.G. Bohlen, W. Von Oertzen et al., Nucl. Phys. A {\bf 180}, 385-401 (1972).
\bibitem{ref.19} R. Tripathi, R. Tripathi, K. Sudarshan et al., Phys. Rev. C {\bf 79}, 064604 (2009).
\bibitem{ref.20} A. Gamp, W. Von Oertzen, H. G  Bohlen et al., Zeitschrift fur Physik, {\bf 261}, 283-304 (1973).
\bibitem{ref.21} G. S. Mallick, S. K. Agarwalla, B. Sahu, C. S. Shastry, Phys. Rev. C {\bf 73}, 054606 (2006).
\bibitem{ref.22} B. Sahu, G. S. Mallick and S. K. Agarwalla, Nucl. Phys. A {\bf 727}, 299 (2003).
\bibitem{ref.23} Joseph N. Ginocchio, Ann. Phys. (N.Y.) {\bf 152}, issue 1: 203-219 (1984).
\bibitem{ref.24} D.J. Hinde, A.C. Berriman, M. Dasgupta et al., Phys. Rev. C {\bf 60}, 054602 (1999).
\bibitem{ref.25} B. B. Back, R. R. Betts, J. E. Gindler et al., Physical Review, C {\bf 32}, (1985) 195.
\bibitem{ref.26} K. E. Rehm, H. Esbensen, C. L. Jiang et al., Physical Review Letters, {\bf 81}, (1998) 3341.
\bibitem{ref.27} Zhang Huanqiao, Liu Zuhua, Xu Jincheng et al., Nuclear Physics, A {\bf 512}, (1990) 531.
\bibitem{ref.28} A. B. Quint, W. Reisdorf, K.-H. Schmidt, et al., Zeitschrift für Physik A {\bf 346}, (1993) 119.
\bibitem{ref.29} K. -H. Schmidt and W. Morawek, Rep. Prog. Phys.{\bf 54}, 949 (1991).
\bibitem{ref.30} Yu. Ts. Oganessian, A. Yu. Lavrentev, A. G. Popeko, et al., JINR FLNR Scientific Report 1995-1996. Heavy Ion Physics, B. I. Pustylnik (ed.), p. 62 (JINR, E7-97-206, Dubna (Russia), 1997).
\bibitem{ref.31} Yu.Ts.Oganessian, V.K.Utyonkov, Yu.V.Lobanov et al., Phys. Rev. C {\bf 64}, 054606 (2001).
\bibitem{ref.32} K. Satou, H. Ikezoe, S. Mitsuoka, et al., Phys. Rev. C {\bf 65}, 054602 (2002).
\bibitem{ref.33} K. K. Jena, S. Senapati, B. B. Sahu, J. K. Nayak and S. K. Agarwalla, arXiv:2201.03805.
\bibitem{ref.34} K. K. Jena, S. K. Agarwalla, B. B. Sahu, Acta Phys. Pol.B, {\bf 53}, 10-A1 (2022).
\bibitem{ref.35} Kamala Kanta Jena, Santosh Kumar Agarwalla, Bidhubhusan Sahu, New J. Phys. {\bf 25},
 033012 (2023).
\bibitem{ref.36} D. R. Tilley, H. R. Weller, C. M. Cheves, and R. M. Chasteler, Nucl. Phys. A {\bf 595}, 1 (1995).
\bibitem{ref.37} National Nuclear Data Center, BNL, Upton, NY 11973-5000, https://www.nndc.bnl.gov
\bibitem{ref.38} C. Simenel, D.J. Hinde, R. du Rietz et al., Phys. Lett. B {\bf 7101}, 607 (2012).
\bibitem{ref.39} D..J. Hinde, M. Dasgupta, and A. Mukherjee, Phy. Rev. Lett. {\bf 89}, 282701 (2002).
\bibitem{ref.40} J. R. Birkelund and J. R. Huizenga, Annu. Rev. Nucl. Part. Sci. {\bf 33}, 265 (1983).
\bibitem{ref.41} S. G. Steadman and M. J. Rhoades-Brown, Annu. Rev. Nucl. Part. Sci. {\bf 36}, 649 (1986).
\bibitem{ref.42} Y. Eisen and Z. Vager, Nucl. Phys. A {\bf 187}, 219 (1972).
\bibitem{ref.43} B. Sahu, L. Satpathy, and C. S. Shastry, Phys. Lett. A {\bf 303}, 105 (2002)
\bibitem{ref.44} B. Sahu, S. K. Agarwalla, and C. S. Shastry, Nucl. Phys. A {\bf 713}, 45 (2003).
\bibitem{ref.45} B. Sahu et. al., Phys. Rev. C {\bf 77}, 024604 (2008). 
\bibitem{ref.46} R. R. Swain et. al., Int. J. Mod. Phys. E {\bf 29}, 2050016 (2020).
\bibitem{ref.47} Hiroshi Ikezoe, Kenichirou Satou et al., Progress of Theoretical Physics 
Supplement, No. {\bf 154}, 45 (2004).





\end{thebibliography}
\end{document}